\documentclass[preprint2]{aastex}
\shorttitle{X-ray Cycles}

\begin{document}
\bibliographystyle{apj}

\title{Constraints on the ubiquity of coronal X-ray cycles\thanks{Based on observations obtained with {\it XMM-Newton}, an ESA science mission with instruments and contributions directly funded by ESA Member States and NASA.}}

\author{John Hoffman}
\affil{Department of Physics, University of Illinois at Urbana--Champaign, 1110 West Green Street, Urbana, IL 61801}
\email{hoffma24@illinois.edu}

\author{Hans Moritz G\"{u}nther, Nicholas J. Wright}
\affil{Harvard-Smithsonian Center for Astrophysics, 60 Garden Street, Cambridge, MA 02138}
\email{guenther@head.cfa.harvard.edu, nwright@cfa.harvard.edu}

\begin{abstract}
Stellar activity cycles are known to be a widespread phenomenon amongst moderately active solar-- and late--type stars from long-term periodic variations in chromospheric Ca {\sc ii} H and K emission lines, yet to date only a handful of coronal X-ray cycles are known. We have surveyed serendipitously observed stellar sources in fields observed multiple times in the last decade by {\it XMM-Newton} and present our analysis of 9 stars from 6 fields. Since our sample is flux--limited, it is strongly biased towards higher levels of X-ray activity. We fit a single temperature {\sc apec} spectrum to each source and search for significant periodicities using a Lomb--Scargle Periodogram (LSP). We use a Monte Carlo (MC) algorithm to yield robust analysis of the statistical significance of cycle detections and non--detections. None of the 9 stellar lightcurves show any convincing indications of periodicity. From MC simulations we simulate the detection capabilities of our methodology and, assuming a uniform distribution of cycle periods and strengths over the domain searched, we conclude with 95\% confidence that less than 72\% of the stars represented by our sample of active stars have 5-13 year coronal X-ray cycles. 
\end{abstract}

\keywords{Stars: activity, stars: coronae, X-rays: stars}
\maketitle
\section{Introduction}

The 11 year sun--spot cycle was discovered over 160 years ago by \citet{1843AN.....20..283S}. Later, the subsequent discovery by \citet{1908ApJ....28..315H} that the global solar magnetic field reverses polarity after each sun--spot cycle led to our current understanding of the 11 year solar activity cycle as arising out of an underlying 22 year magnetic cycle. 

In 1966, in order to ascertain whether or not other stars exhibited long--term periodic behavior, O. C. Wilson started a long--term monitoring program of chromospheric activity in solar--type main sequence stars. He defined the S--index, which characterizes the strength of the Ca {\sc ii} H and K lines relative to the photospheric continuum \citep{1968ApJ...153..221W}. Through monitoring a wide range of stars in the chromospheric \ion{Ca}{2} H and K emission lines via the S--index, we now understand that the activity cycle of our Sun is a more general phenomenon amongst solar type stars \citep{1978ApJ...226..379W,1995ApJ...438..269B}. What factors determine the durations of these cycles, however, remains elusive to theoretical explanation.

The Ca {\sc ii} H and K lines are well established indicators of magnetic activity. Additionally, virtually all solar-- and late--type main sequence stars, including the Sun, have hot coronae that emit in the X-ray bands \citep{1981ApJ...245..163V}. The energy required to heat the corona is supplied by the stellar magnetic field, the dynamics of which are governed by an $\alpha-\Omega$ dynamo involving differential rotation and turbulence in the convective envelope \citep[e.g.][and references therein]{1985ARA&A..23..413R,2003A&ARv..11..287O}, which is clearly manifested in the stellar activity---rotation relationship: stars that rotate faster are typically more X-ray luminous \citep[e.g.,][]{1981ApJ...248..279P,0004-637X-743-1-48}.

Early efforts to statistically determine a trend of long--term variability in X-ray stars from {\it Einstein} and ROSAT data are summarized in  \citet{1998ASPC..154..223S}. Stars that are X-ray faint typically show low levels of short-timescale variability \citep{2004ApJ...611.1107F,2010ApJ...725..480W}, but may vary on longer timescales. \citet{2003A&A...404..637M} compare the solar variability to stellar data and find that this is consistent with cycles explaining much of the apparent luminosity spread in solar-like\footnote{``solar-like" refers to spectral type (F--M), and not levels of X-ray activity.} stars. Stars that are X-ray bright ($L_X>10^{28}$ erg~s$^{-1}$) are not believed to exhibit long term X-ray cycles and instead have X-ray luminosities that vary erratically on short timescales \citep[see review by][]{1998ASPC..154..223S}. In this paper, we add experimental credence to this assertion by demonstrating an absence of long-term cycles in X-ray active solar-like stars.
 

Currently, long term X-ray cycles have been found in only a handful of stellar sources: \object{61 Cygni A} \citep{2006A&A...460..261H}, \object{HD~81809} \citep{2008A&A...490.1121F}, \object{$\alpha$ Centauri B} \citep{2005A&A...442..315R,2008ApJ...678L.121A,2009ApJ...696.1931A} and our Sun. The amplitudes of the stellar X-ray cycles in these objects are typically larger than their respective S-index cycles by up to an order of magnitude, as is the case in the Sun \citep{2003ApJ...593..534J}. 

The large levels of periodic X-ray variability expected from solar-like main-sequence stars mean that their cycles should be detectable with modern X-ray observatories such as {\it XMM-Newton} that have been in service for over a decade. In this article we analyze serendipitous stellar sources detected in {\it XMM-Newton} fields that have been frequently observed over the lifespan of the instrument ($\sim$12 years) using publicly available data from the {\it XMM-Newton} archive. In Section~\ref{sect:obs} we discuss the observations used for this study and the methods employed. In Section~\ref{sect:results} we present the observational results and Monte Carlo (MC) simulations to quantify our detection limits. In Section~\ref{sect:discussion} we discuss the implications of these results on our understanding of stellar X-ray emission and long-term, solar-like X-ray cycles in other stars. We end with a presentation of our conclusions in Section~\ref{sect:conclusions}.

\section{Observations and Data Analysis}
\label{sect:obs}
In {\it XMM-Newton}'s nearly 12 year history, numerous fields have been observed semi-regularly for purposes ranging from calibration (e.g. Capella) to measuring long term periodic variability in X-ray luminosities (e.g. HD~81809). These fields offer a rare opportunity for the detection of long term stellar X-ray cycles, which are believed to correlate strongly with chromospheric activity cycles. While central sources have been analyzed \citep[e.g., HD~81809,][]{2008A&A...490.1121F}, serendipitous sources have not, to our knowledge, been checked for periodic behavior. Thus, for fields which have a long and consistent history of {\it XMM-Newton} observations, we check for periodicities in the lightcurves of serendipitous stellar sources. 

\subsection{Field selection}

Target fields that have been frequently observed (i.e. that have a baseline of at least six years and have been observed during at least 5 semesters between 1999 and 2012) were selected for analysis. We will show below that all candidate sources are relatively faint. To provide enough counts for a stable fit of the spectrum, we only use observations with a Good Time Interval (GTI) of $\geq$10~ks for one or more of the MOS1, MOS2, and PN cameras. Any data mode is acceptable. For instance, even though modes \texttt{PPW2} and \texttt{PPW4} have small central windows that do not read out parts of the field near the central source, these exposures still contain most serendipitous sources in the field. Six fields pass these selection criteria. Table~\ref{tab:fieldinfo1} summarizes the observation timing and baselines for the 6 fields used in this paper. ``Distinct observations" are the number of half-year intervals between 1999 and 2012 containing at least one usable observation.
\begin{table}
\centering
\caption{Field Characteristics\label{tab:fieldinfo1}}
\begin{tabular}{llll}
\hline \hline
Central & Abbr. &  Baseline & Distinct \\
 source & name &  (yr) & obs. \\
\hline
Capella & CAP & 7.48 & 6 \\
Zeta Puppis & ZP & 9.98 & 13 \\
PKS 2155-304 & PKS & 10.9 & 16  \\
3C273 & 3C & 10.49 & 11	\\
HD~81809 & HD & 6.0 & 5 \\
Mkn 421 & MK & 10.98 & 7 \\
\hline
\\
\end{tabular}
\end{table}

\subsection{Source selection and identification of stars}

X-ray sources are selected by cross-matching the 2XMMi Serendipitous Source Catalogue \citep{2009A&A...493..339W} with the optical NOMAD catalog \citep{2004AAS...205.4815Z}, using a matching radius of 5\arcsec consistent with the \emph{XMM-Newton} PSF for off-axis sources. For bright central sources, the 2XMMi catalog contains many spurious detections in the vicinity that are related to read--out streaks in the original data. These read--out streaks are manually removed from our target list. 

We impose a minimum flux limit of 25 counts in a 10~ks observation, which equates to a maximum flux uncertainty of 20\% from Poisson statistics. For an observation of 10~ks and 25 counts this equates to a flux limit of $1.4\times10^{-14}$~erg~cm$^{-2}$~s$^{-1}$ for a typical stellar observation and stellar properties (solar abundance, plasma temperature of 1~keV and observations using the medium filter for the MOS).

A number of steps are taken to identify stellar X-ray sources. First, because stars have relatively soft X-ray spectra compared to most extragalactic sources, we require a hardness ratio of HR1~$>0$ \citep[see][]{2009A&A...493..339W}. We then search the literature for available photometry and spectroscopy to allow us to classify sources as stellar or non-stellar. Due to the large number of non-stellar sources in deep X-ray observations, we require a high-confidence verification of the stellar nature of our candidates, discarding all sources that lack sufficient data to confidently classify them. We use three methods for identifying stellar sources based on the availability of certain observations and measurements, which, in order of priority, are:

\begin{itemize}
\item {\bf Spectra} showing that the source is stellar in nature. Available for five targets (see Table~\ref{smtb}).
\item {\bf Proper motions} showing that the source is sufficiently nearby as to have a measurable motion on the plane of the sky with a significance of 3$\sigma$ from the catalog of \citet{2008A&A...488..401R}. For stars with proper motion, but no spectrum, we require that the photometry --if available -- is also consistent with stellar-like colors \citep{2007AJ....134.2398C}. Three targets meet these criteria (\#4, \#5, \#7).
\item {\bf Photometry} indicating stellar-like colors \citep{2007AJ....134.2398C} in all bands. One target meets this criterion (\#3).
\end{itemize}

In addition to these requirements, stellar targets without available spectral information are required not to have been identified as extragalactic by any other studies incorporating observations at other wavelengths, morphology, or more complex multi-wavelength diagnostics. These criteria reduce the initial list of 72 sources to a high-confidence list of 9 stellar sources (see Table~\ref{smtb}).

\subsection{Spectral types and distances}

For these 9 stellar sources, spectral types are determined from the best available data, either via spectroscopy (for 5 sources) or from the photometric color with the largest baseline (for 4 sources) and using the empirical data of \citet{2012ApJ...746..154P}. All sources are assumed to lie on the main-sequence, a reasonable assumption since our targeted fields are away from known star forming regions and X-ray emission from late-type post-main sequence stars is rare and weak \citep{1979ApJ...229L..27L}. The determined spectral types range from F-type to early M-type. Extinctions were calculated based either on the available photometry (for stars with known spectral types), or from an initial estimate of the distance (from the reddened magnitudes) and assuming an extinction of $A_V = 1$ mag / kpc, up to a maximum extinction of the total integrated Galactic extinction in that line-of-sight from \citet{schl98}. Distances were determined using 2MASS \citep{2006AJ....131.1163S} $K_s$-band photometry and the empirical absolute magnitudes from \citet{2012ApJ...746..154P}. For one source, Gliese~195~A, a parallax is available \citep{1952QB813.J45......} and is therefore used to determine its distance. The determined stellar distances range from 14 to 663~pc.

\subsection{Data analysis}

Observational data are obtained from the {\it XMM-Newton} archive and analyzed using SAS \citep[version 11.0.0,][]{2006IAUSS...6E..13G}. We apply all standard selection criteria for rejecting cosmic rays and filtering out the background rate to avoid contamination from solar proton flares. For all sources, spectra are extracted and the appropriate response files are generated. Fitting of spectral models is performed in Sherpa \citep{2011ascl.soft07005F}.

Due to the rather large point spread function of {\it XMM-Newton} and the relatively dim nature of the selected sources, a 450-500 pixel ($\sim$8-9$^{\prime\prime}$) radius is used for the source region and a 2500 pixel ($\sim$46$^{\prime\prime}$) radius is used for the background region, manually chosen to avoid other 2XMMi sources. One background region is chosen for each field to simplify the processing.\footnote{Using background regions from different chips yielded consistent results in a small test run.}

\subsection{Fitting {\sc apec} models}

A single temperature {\sc apec} model is fitted to extracted spectra that contain a minimum of 125 counts within the 0.4-7 keV range after subtracting background counts. We fix the interstellar absorption using the $A_V$ values from table~\ref{smtb} and the relation $N_H = 1.8\times 10^{21} \frac{1}{\textnormal{mag cm}^2} A_V$ \citep{1995A&A...293..889P}. We employ the Levenberg-Marquardt (LM) fitting algorithm \citep[][called \texttt{levmar} in \texttt{sherpa}]{levmar} with a Gehrel's $\chi^2$ statistic \citep{1986ApJ...303..336G} to fit {\sc apec} models to the observed spectra. If the Gehrel's $\chi^2$ value of a LM fit exceeds a threshold value heuristically set to 0.7, we seek an improvement of the fit. First, the LM algorithm is again tried with five arbitrary and distinct initial parameter values. If the Gehrel's $\chi^2$ of the fit is still above the heuristic threshold, the fitting method is changed to the more accurate but slower differential evolution method \citep{moncar} (called \texttt{moncar} in \texttt{sherpa}) as a second effort for improving the fit. 

At the end of the fitting procedure, if the Gehrel's $\chi^2$ is below 0.25 or above 3, 1$\sigma$ flux uncertainties are not computed. A $\chi^2$ above 3 strongly suggests that the model does not adequately fit the data. A $\chi^2$ below $\sim$0.25 suggests that the {\sc apec} model fit is not well constrained. Flux values without estimated 1$\sigma$ uncertainties are not used in later analysis.

Unless otherwise stated, X-rays are defined to be in the 0.4-7 keV band. $F_X$ is the integrated energy flux over this range. $L_X$ is the unabsorbed luminosity over this range. The $F_X$ values are obtained from integrating the fitted {\sc apec} model. Fit errors for the integrated $F_X$ are calculated from the uncertainty of the model normalization.
	
\section{Results}
\label{sect:results}
\subsection{The stellar sample}

Table~\ref{smtb} summarizes the properties of our stellar sample. Uncertainties in $L_X$ and $L_X/L_{bol}$ are about 0.2~dex, including the statistical uncertainty from the fitted X-ray model and the uncertainty in distance from the photometry. The spectral types of our sample range from F to early M, the activity level ($\log L_X/L_{bol}$) from $-3$ to $-6$. Active stars are known to saturate at an activity level of $\log(L_X/L_{bol})\sim-3$ , while the activity level drops for older or more slowly rotating stars \citep{1994ApJS...91..625S}. Naturally, stars as inactive as our Sun are not part of this serendipitous sample because they are too X-ray faint.

\begin{table*}
\caption{Summary of stellar sources}
\begin{tabular}{llccccccccc}
\hline
\#	& Field &	RA		& Dec		& Spectral		& $V$		& $K_s$	& $A_V$	&	d	& $\log(L_X)$	& $ \log(L_x/L_{bol})$ \\
\hline
& name & (h:m:s)		& (d:m:s)		&	type		& (mag)		& (mag)	& (mag)		& (pc)		& ($erg/s$)	&				\\
\hline
1$^a$ & CAP	& 05:16:32.0 & 46:08:27.6 & F0$^d$		& 8.06		& 7.23	& 0.11		& 123		&	$28.6$ & $-5.8$	\\ 
2$^b$ & CAP	& 05:17:24.0 & 45:50:20.4 & M1$^e$	& 10.16		& 5.95	& 0.13		& 13.9$^i$	&	$28.2$ & $-4.0$	\\
3     & ZP	& 08:04:22.6	& -39:50:28.7	& G0			& 13.37		& 11.57	& 0.43		& 419		&       $30.3$ & $-3.6$			 \\
4     & HD	& 09:27:12.3	& -05:57:16.9	& K3			& 15.46		& 13.07	& 0.11		& 572		&	$30.5$ & $-2.6$		 \\
5     & HD	& 09:28:34.3	& -06:03:45.7	& F9			& 13.37		& 11.91	& 0.11		& 663		&       $30.3$ & $-3.9$			 \\
6$^c$ & MK & 11:04:43.8	& 38:14:48.5	& F5$^f$		& 7.51		& 6.39	& 0.05		& 64.9		&	$29.0$ & $-5.1$	\\	
7     & 3C	& 12:28:28.6	& 01:54:49.3	& K0			& 12.68		& 10.67	& 0.06		& 240		&	$29.0$ & $-4.4$		 \\
8     & 3C	& 12:28:37.2	& 01:57:20.9	& M1$^h$		& 14.09		& 10.04	& 0.06		& 79.5		&	$28.8$ & $-3.4$		  \\
9     & 3C	& 12:29:42.4	& 01:55:25.3	& G0$^g$		& 12.88		& 10.89	& 0.06		& 383		&	$30.3$ & $-3.5$		 \\
\hline
\end{tabular}
\newline
{\it Other names:} $^a$BD+45 1076; $^b$Capella H, G~96-29, Gliese 195 A; $^c$HD 95976 \\
{\it Notes:} Spectral types extracted from the literature where a citation is provided, or from available colors. $V$ magnitudes taken from \citet{hog2000} and the NOMAD catalog \citep{2004AAS...205.4815Z}. $K_s$ magnitudes extracted from the 2MASS catalog \citep{2006AJ....131.1163S} and distances determined photometrically from this (with the exception of \#2, which has a parallax). \\
{\it References:}
$^d$\citet{1975ascp.book.....H}, 
$^e$\citet{2001MNRAS.328...45M},
$^f$\citet{1918AnHar..91....1C}, 
$^g$\citet{refId0}, 
$^h$\citet{2008ApJS..178..339C}, 
$^i$\citet{1952QB813.J45......}. 
\label{smtb}
\end{table*}

\subsection{The Lomb--Scargle Periodogram}

After fitting a single temperature {\sc apec} model to the spectra of all 9 stellar sources and integrating the fitted model over the 0.4-7 keV range to obtain flux estimates for each exposure, the resulting lightcurve is analyzed for the presence of periodic behavior. This is done using a Lomb--Scargle Periodogram (LSP) \citep{lomb, 1982ApJ...263..835S}. Derived quantities from the LSP are (1) the frequency of the best fit sinusoidal function to the data (known as the peak frequency) and (2) the ``False Alarm Probability" (FAP) of the strongest periodicity found. The FAP is defined to be the probability that periodic behavior of a certain strength could be found in Gaussian noise. 

\subsection{Cycle detection}

\citet {1995ApJ...438..269B} consider a FAP $\le$ 0.001 ($\ge$99.9\% confidence) as a detection of periodicity in their analysis of S-index data from the Mount Wilson Observatory. In our case, X-ray flux uncertainties are large and most stars have been observed sparsely in the last 12 years. Thus, we do not expect the lightcurves of the stellar sources in this paper to produce such high--confidence statistics. 

Instead of employing a simple FAP cutoff, we use a MC algorithm to assess the statistical likelihood of LSP results by taking into consideration the flux uncertainties, which vary from observations to observation. We consider a lightcurve to be periodic (or at least warrant further investigation) if 68\% of all MC--sampled lightcurves have a FAP below 0.30.

\subsection{Monte Carlo simulations}

While periodogram analysis offers an invaluable statistical method for determining the presence of periodicities, the LSP calculation itself does not account for uncertainties in the data. Therefore, we use a MC algorithm to determine the statistical significance of a detected periodicity. 

\begin{figure*}
\centering
\includegraphics[width=\linewidth]{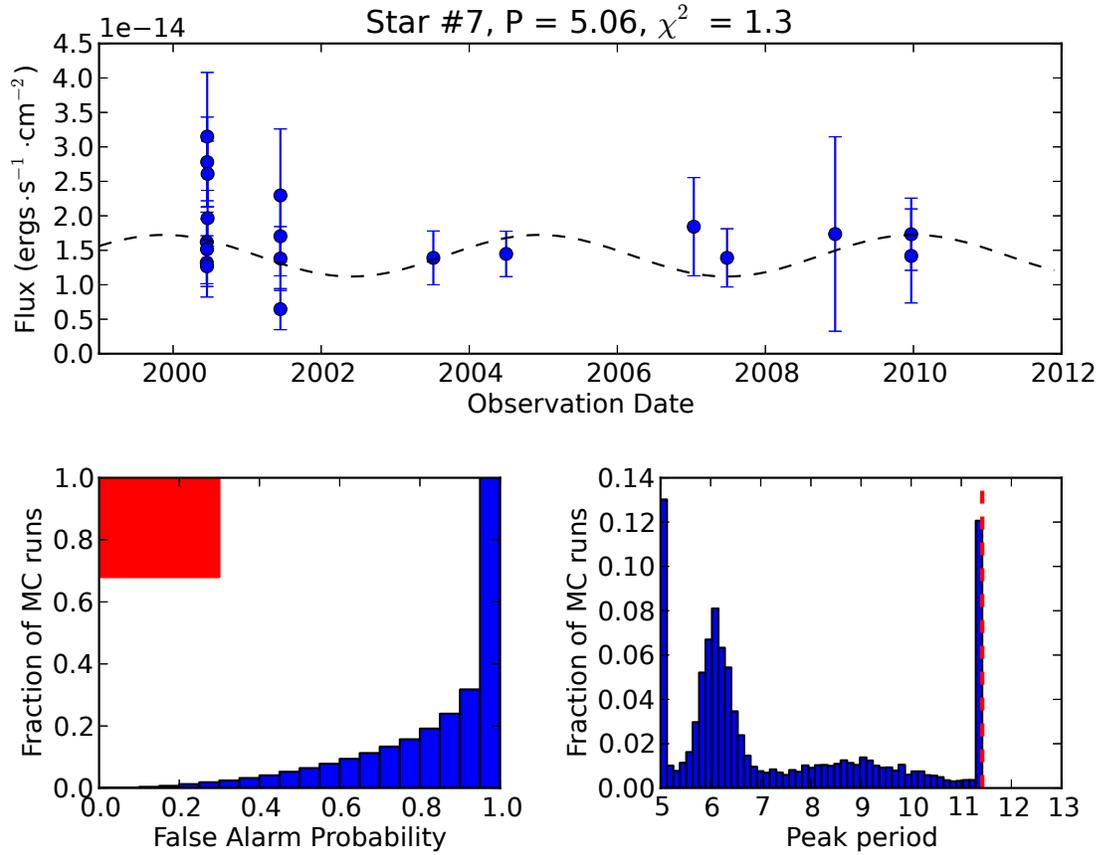}
\caption{\label{3ca3pf} {\it Top:} Lightcurve for \#7. The dashed line is the best--fit sinusoid (of period $P_m$, the most likely peak--period according to MC analysis) through the data. {\it Bottom Left:} A cumulative histogram of FAP values from the $N=1000$ Monte Carlo--sampled lightcurves from \#7. If the cumulative histogram of FAP values falls within the square in the upper left corner, the lightcurve is periodic by the standards employed in this paper. {\it Bottom Right:} A non--cumulative histogram of peak periods (inverse of the Lomb--Scargle peak frequency) detected in the MC--sampled lightcurves of \#7. The vertical dashed line indicates the maximum period investigated by the Lomb--Scargle Periodogram, $1.2\times B$ where $B$ is the baseline for usable observations of \#7.}
\end{figure*}

In a lightcurve containing $T$ data points with uncertainties, the MC algorithm samples $T$ flux values in the following way for each step ($N=1000$ total steps) of the simulation:
$$F_{MC}(t_i) = F_i + X_i \; ,$$
where $F_i$ is the $i$th observed flux (at time $t_i$) and $X_i$ is a random variable sampled from a Gaussian distribution of mean 0 and a standard deviation equal to the uncertainty of the empirical flux value. Figure \ref{3ca3pf} shows the lightcurve, a cumulative histogram of FAP values and a histogram of the strongest periods (inverse of the peak frequency in the LSP) detected in \#7, a K1 star in the 3C 273 field.

\subsection{The test case: HD~81809}

\begin{figure}
\centering
\includegraphics[width=\linewidth]{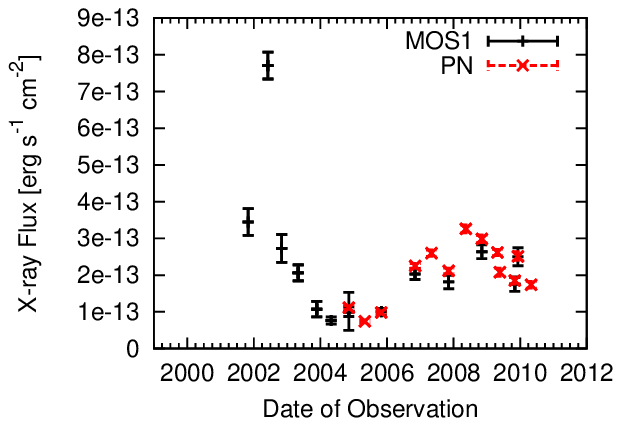}
\includegraphics[width=\linewidth]{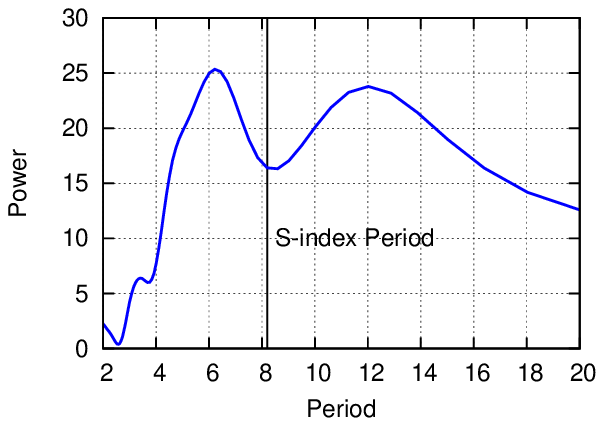}
\caption{{\it Top:} A plot of fitted {\sc apec} flux for HD~81809, showing a strong $\sim8$ year periodicity. For better clarity, MOS2 data is not shown here but agrees with MOS1 data within 1$\sigma$. {\it Bottom:} LSP of empirical flux for HD~81809. An S-index cycle period of 8.2 years was found by \citet{1995ApJ...438..269B}. The apparent disagreement between the LSP peak frequency for the X-ray lightcurve of HD 81809 and the 8.2 year S-index period is likely due to semi--regular observation timing.}
\label{fig:hd81809lc}
\end{figure}

As a test case, we present an analysis of a stellar source with a known X-ray cycle. HD~81809 has an 8.2 year S-index cycle \citep{1995ApJ...438..269B} and an X-ray cycle that appears in phase with the S-index cycle \citep{2008A&A...490.1121F}. We use all publically available {\it XMM-Newton} data for HD~81809 and carefully determine the presence and significance of any periodicities in its lightcurve. Figure \ref{fig:hd81809lc} displays fitted {\sc apec} flux values as well as a LSP. Indeed, the periodic behavior has continued with what appears to be an $\sim$8 year period. 

The LSP produces a local minimum at the expected period surrounded by two peaks at a period of $~6.2$ years and $~12$ years. The MC simulations confirm that the 6~yr peak has a FAP of $<$5\% in 100\% of the MC steps. The MC simulations also show that the shift in period compared to the S-index observations is an artifact caused by infrequent time-sampling of the \emph{XMM-Newton} observations compared to the observations of \citet{1995ApJ...438..269B}. Similar behavior can be found in Fourier analysis of X-ray data for 61 Cyg, where the double peak in the Fourier power close to the known S-index period is revealed to be an artifact of infrequent and semi-regular observation times: A similar phenomenon occurs in the Fourier analysis of S-index data when sampled at the same times as X-ray observations \citep{2006A&A...460..261H}.

\subsection{Detection capabilities}

Despite the inaccuracy of the exact cycle period, analysis of HD~81809 shows that our method for determining the \emph{presence} of stellar cycles is statistically sound. No significant periodic behavior was found in the lightcurves of any of the 9 stars in our sample.

\begin{figure}
\centering
\includegraphics[width=\linewidth]{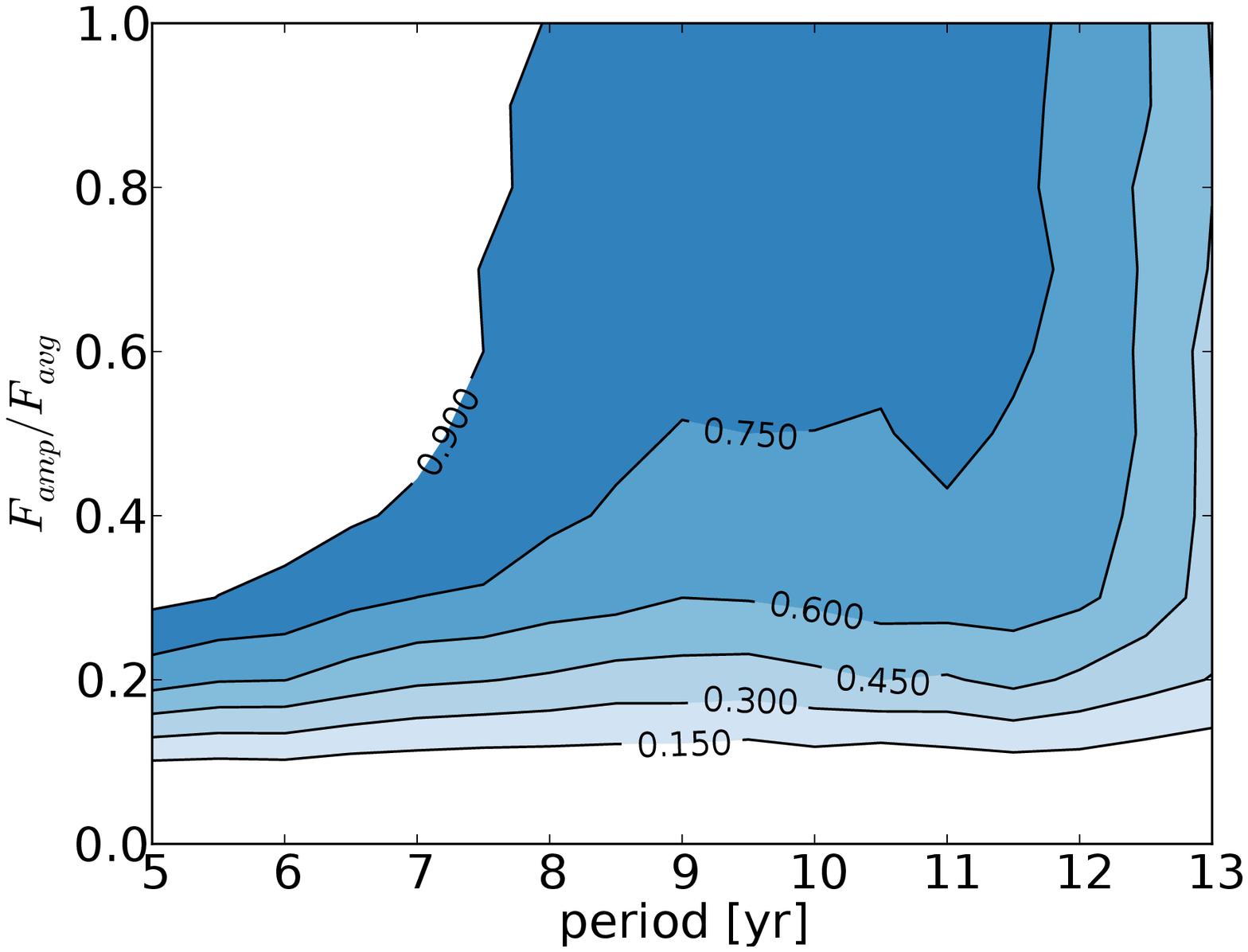}
\includegraphics[width=\linewidth]{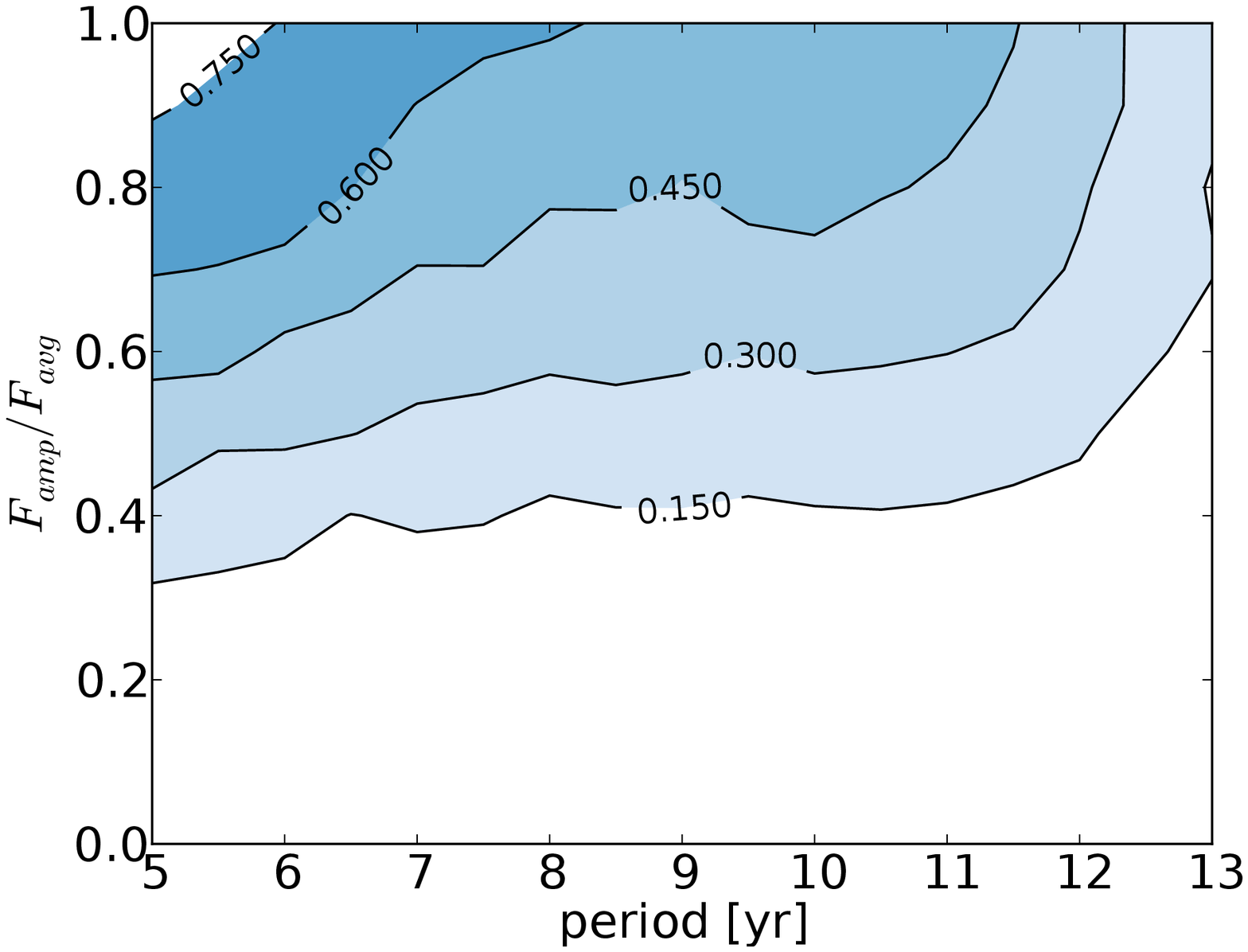}
\caption{Contour plots of the probability of observing an underlying cycle with a Lomb--Scargle false alarm probability less than 0.3 for a given $F_{amp}/F_{avg}$ and cycle period. This is calculated assuming that observations are randomly distributed over a baseline of 11 years with Gaussian flux uncertainties of $\sigma\times F_{avg}$ for each observation. \emph{top:} 20 observations, small flux uncertainties ($\sigma=0.15$); \emph{bottom:} 10 observations, $\sigma = 0.3$.}
\label{fig:cont}
\end{figure}

To determine which cycles our method is capable of detecting, we use MC simulations again, assuming that a source's X-ray flux over time is:

\begin{equation}
f(t) = F_{avg}+F_{amp}\sin{\left(\frac{2\pi t}{P}+\delta\right)}
\label{eqn:ft}
\end{equation}

\noindent
and that the observations are subject to Gaussian uncertainty with standard deviation, $\sigma$. Thus, for a given period $P$ and cycle amplitude relative to the average flux ($F_{amp}/F_{avg}) = s$, a MC simulation creates $T$ randomly spaced data points over a span of $S$ years, produces a LSP and calculates the FAP of the best fit sine function. $N$ sampled lightcurves are created for each ($P$, $s$) value. We then calculate the fraction of $N=1000$ MC lightcurves that produce a FAP less than 0.3. Several contour plots are shown in Figure \ref{fig:cont}.

Contour plots like those in Figure \ref{fig:cont} are calculated via a similar MC algorithm for each individual source. Instead of sampling $N$ observations randomly distributed over the baseline with a constant uncertainty, the observation dates and flux uncertainties for the particular star of interest are used to generate sample lightcurves in the MC algorithm. 

\section{Discussion}
\label{sect:discussion}

The time span covered by the observations differs for each analyzed field and ranges between 6 and 11 years. Most sources have more than 30 exposures with a usable time longer than 10~ks. However, several exposures are often taken within a few days and thus do not contribute more information to the periodogram. Typically, about 10 datapoints with sufficient spacing exist. The statistical errors depend on the source flux and the exposure time. Source flux varies from source to source, and exposure time varies even between exposures. Here, a typical value of relative error on the flux is about 25\%. 

We now seek to use our results to put an upper limit on the prevalence of coronal X--ray cycles in stars of spectral type F through M that have 5--13 year periods. We calculate $P(FAP<0.3)>0.68$ contours for each source and use these to create a binary function over the $P$--$s$ domain that is equal to 1 when $P(FAP<0.3)>0.68$ and 0 otherwise. The binary functions for each star are summed to generate a map ($Z(P,s)$) that represents the number of cycles that \emph{could} have been detected using our method for a cycle of particular strength and period. 

We assume for simplicity that the intrinsic rate of coronal X-ray cycles is approximately constant across spectral types F--M. This assumption may or may not be justified: the S--index monitoring presented in \citet{1995ApJ...438..269B} did not investigate M stars (except for one for which no cycle was found), however, monitoring by \citet{2012arXiv1204.4101M} has detected possible S--index cycles in M--dwarfs which fall within our 5--13 year domain. \citet{1995ApJ...438..269B} do see differences in chromospheric cycles between F, G and K type stars, such that e.g. almost all K-type stars with a low S--index have a cycle, whereas F--type stars with low S--index typically have flat activity curves. However, the weakness  of the chromosphere and the low contrast of the photosphere make the detection of cycles on F--type stars more difficult, so that it is not clear if the cycle frequency actually varies with spectral type.

\begin{figure}
\centering
\includegraphics[width=\linewidth]{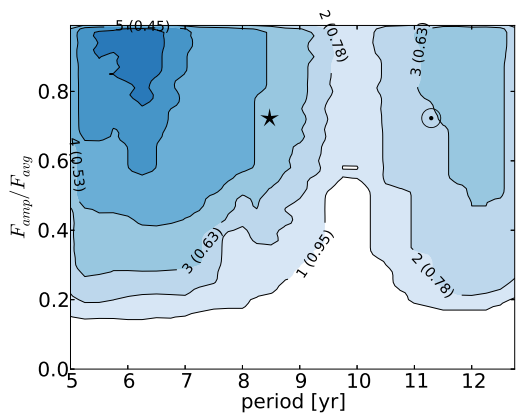}
\caption{\label{2sigup} Contour plot of cycles that could be detected using our methodology. The $z$--axis represents the number of stars in our sample for which a certain type of cycle should have been detected, based on MC simulations. 2$\sigma$ upper limits on the intrinsic rate ($R$) of coronal X-ray cycles in F--M stars, assuming that $R$ is relatively constant across these spectral types, are provided in parentheses. The relative amplitude and period of X-ray cycles for both the Sun and HD 81809 are denoted by solar and star symbols respectively. The plot has been slightly smoothed for display.}
\end{figure}

For a particular kind of cycle $(P_c,s_c)$, $Z(P_c,s_c)$ is then the effective sample size. The probability of \emph{not} detecting a $(P_c,s_c)$--type cycle is $(1-R)^{Z(P_c,s_c)}$, where $R$ is the intrinsic rate of $(P_c,s_c)$--type coronal X-ray cycles for F--M stars. A 2$\sigma$ upper--limit on $R$ would imply that $R\le R_{2\sigma}$ where $R_{2\sigma}$ is defined to be such that the probability of not detecting a certain type of cycle, $(1-R_{2\sigma})^{Z(P_c,s_c)}$, is 0.05. Solving for $R_{2\sigma}$ in terms of $Z$, $R_{2\sigma}(P,s) = 1-(0.05)^{1/Z}$. A plot of $Z$ is provided in Figure \ref{2sigup}, with implied $R_{2\sigma}$ values provided in parentheses.

While there is a notable area where $R_0$ is high (i.e. where the 2$\sigma$ upper limits on the underlying rate of coronal X-ray cycles is weak), the statistics from this sample strongly limit the prevalence of short term cycles (5--8 year periods) in our sample. Since both the sun and HD 81809 have $s$ values close to 0.7, we can say with 95\% confidence that fewer than $\sim$50\% of the stars in our sample have 5--8 year X--ray cycles with amplitudes of $s = 0.7$. 

In addition, if one makes the assumption that coronal X-ray cycles should have strengths and periods that are more or less uniformly distributed over the ($0\le s\le1$) and (5 years $\le P \le 13$ years) domain, Figure \ref{2sigup} implies a 2$\sigma$ upper limit on the ubiquity of coronal X-ray cycles in stars represented by our sample. Under this assumption, our statistics imply that less than $\sim72\%$ of those stars have coronal X-ray cycles.

All statistical constraints given thus far are valid for a stellar population that is well represented by our sample, i.e. the population of stars detected in a serendipitous X-ray survey. Since we are limited by the flux of sources in the field, our sample is biased towards X-ray active stars (Malmquist-bias). Only two sources have activity levels $L_X/L_{bol} < -4.5$ and even the least active one is still more active than the Sun by an order of magnitude or more. Thus, we cannot draw reliable conclusions about F--M stars in general from our data. This requires a volume limited sample as e.g. in \citet{2004A&A...417..651S}.

The lack of detection of activity cycles in our sample would therefore support the view that in active stars coronal X-ray cycles are not common. This was observed in the S-index monitoring by \citet{1995ApJ...438..269B}. In X-rays the absence of cycles in active stars has been suggested by \citet{1998ASPC..154..223S} on the bases of more fragmentary data. As stars age their median luminosity decreases \citep[see reviews of ][ and references therein]{1985ARA&A..23..413R,2003SSRv..108..577F} and they develop coronal cycles with larger amplitudes of variation \citep{2003A&A...404..637M}.

\section{Conclusions}
\label{sect:conclusions}

We have identified 9 stellar X-ray sources that have frequently been observed in repeat {\it XMM-Newton} observations over the lifetime of the observatory. From single temperature {\sc apec} model fits to extracted spectra, we have generated lightcurves of all stellar sources and tested the lightcurves of these stars for the presence of underlying periodicities using a Lomb--Scargle Periodogram assisted by a Monte Carlo algorithm. No periodicities were found. 

Using Monte Carlo simulations, we calculated which kinds of cycles should be detectable with our methodology for each of the 9 stars in our sample. We term the underlying rate of coronal X-ray cycles in F--M stars that have 5--13 year periods as $R$. We assume that $R$ is constant across spectral types F through M and that the short--term X-ray variability of our stellar sources is negligible compared to any long-term periodic behavior. Using these assumptions, we calculate 2$\sigma$ upper limits on $R$ for each type of cycle investigated in this study (cycles of any strength that have 5--13 year periods). Our sample is biased towards stars with higher levels of X-ray activity; the least active star is at least an order of magnitude more X-ray active than our sun. If one makes the assumption that X-ray active solar-type stars have cycles with periods and strengths that are uniformly distributed over the ($0\le s\le1$) and ($5$ years $\le P \le 13$ years) domain, we can conclude with 95\% confidence that less than 72\% of the active main sequence stars have 5--13 year coronal X-ray cycles of any strength. We stress that these numbers do not apply to the stellar population as a whole, which contains a large fraction of stars that have significantly lower levels of X-ray activity.


\begin{acknowledgments}

JH would like to thank Marie Machacek and Jonathan McDowell for their invaluable guidance and assistance with this project throughout the summer of 2011. This work is supported in part by the National Science Foundation Research Experiences for Undergraduates (REU) and Department of Defense Awards to Stimulate and Support Undergraduate Research Experiences
(ASSURE) programs under Grant no. 0754568 and by the Smithsonian Institution. We thank an anonymous referee for his help in improving the paper.

\end{acknowledgments}

\bibliography{shortref}

\begin{thebibliography}{47}
\expandafter\ifx\csname natexlab\endcsname\relax\def\natexlab#1{#1}\fi

\bibitem[{{Ayres}(2009)}]{2009ApJ...696.1931A}
{Ayres}, T.~R. 2009, \apj, 696, 1931

\bibitem[{{Ayres} {et~al.}(2008){Ayres}, {Judge}, {Saar}, \&
  {Schmitt}}]{2008ApJ...678L.121A}
{Ayres}, T.~R., {Judge}, P.~G., {Saar}, S.~H., \& {Schmitt}, J.~H.~M.~M. 2008,
  \apjl, 678, L121

\bibitem[{{Baliunas} {et~al.}(1995){Baliunas}, {Donahue}, {Soon}, \& {et
  al.}}]{1995ApJ...438..269B}
{Baliunas}, S.~L., {Donahue}, R.~A., {Soon}, W.~H., \& {et al.} 1995, \apj,
  438, 269

\bibitem[{{Cannon} \& {Pickering}(1918)}]{1918AnHar..91....1C}
{Cannon}, A.~J. \& {Pickering}, E.~C. 1918, Annals of Harvard College
  Observatory, 91, 1

\bibitem[{{Covey} \& {et al.}(2008)}]{2008ApJS..178..339C}
{Covey}, K.~R. \& {et al.} 2008, \apjs, 178, 339

\bibitem[{{Covey} {et~al.}(2007){Covey}, {Ivezi{\'c}}, {Schlegel}, \& {et
  al.}}]{2007AJ....134.2398C}
{Covey}, K.~R., {Ivezi{\'c}}, {\v Z}., {Schlegel}, D., \& {et al.} 2007, \aj,
  134, 2398

\bibitem[{{Favata} \& {Micela}(2003)}]{2003SSRv..108..577F}
{Favata}, F. \& {Micela}, G. 2003, \ssr, 108, 577

\bibitem[{{Favata} {et~al.}(2008){Favata}, {Micela}, {Orlando}, , \& {et
  al.}}]{2008A&A...490.1121F}
{Favata}, F., {Micela}, G., {Orlando}, S., , \& {et al.} 2008, \aap, 490, 1121

\bibitem[{{Feigelson} {et~al.}(2004){Feigelson}, {Hornschemeier}, {Micela}, \&
  {et al.}}]{2004ApJ...611.1107F}
{Feigelson}, E.~D., {Hornschemeier}, A.~E., {Micela}, G., \& {et al.} 2004,
  \apj, 611, 1107

\bibitem[{{Freeman} {et~al.}(2011){Freeman}, {Nguyen}, {Doe}, \&
  {Siemiginowska}}]{2011ascl.soft07005F}
{Freeman}, P., {Nguyen}, D., {Doe}, S., \& {Siemiginowska}, A. 2011, in
  Astrophysics Source Code Library, record ascl:1107.005, 7005

\bibitem[{{Gabriel} {et~al.}(2006){Gabriel}, {Guainazzi}, {Metcalfe}, \& {et
  al.}}]{2006IAUSS...6E..13G}
{Gabriel}, C., {Guainazzi}, M., {Metcalfe}, L., \& {et al.} 2006, IAU Special
  Session, 6

\bibitem[{{Gehrels}(1986)}]{1986ApJ...303..336G}
{Gehrels}, N. 1986, \apj, 303, 336

\bibitem[{{Hale}(1908)}]{1908ApJ....28..315H}
{Hale}, G.~E. 1908, \apj, 28, 315

\bibitem[{{Heckmann}(1975)}]{1975ascp.book.....H}
{Heckmann}, O. 1975, {AGK 3. Star catalogue of positions and proper motions
  north of -2.5 deg. declination}, ed. {Heckmann, O.}

\bibitem[{{Hempelmann} {et~al.}(2006){Hempelmann}, {Robrade}, {Schmitt}, \& {et
  al.}}]{2006A&A...460..261H}
{Hempelmann}, A., {Robrade}, J., {Schmitt}, J.~H.~M.~M., \& {et al.} 2006,
  \aap, 460, 261

\bibitem[{{H{\o}g} {et~al.}(2000){H{\o}g}, {Fabricius}, {Makarov}, {Urban},
  {Corbin}, {Wycoff}, {Bastian}, {Schwekendiek}, \& {Wicenec}}]{hog2000}
{H{\o}g}, E., {Fabricius}, C., {Makarov}, V.~V., {Urban}, S., {Corbin}, T.,
  {Wycoff}, G., {Bastian}, U., {Schwekendiek}, P., \& {Wicenec}, A. 2000, \aap,
  355, L27

\bibitem[{{J. L\'opez-Santiago} {et~al.}(2007){J. L\'opez-Santiago}, {G.
  Micela}, {S. Sciortino}, \& {et al.}}]{refId0}
{J. L\'opez-Santiago}, {G. Micela}, {S. Sciortino}, \& {et al.} 2007, A\&A,
  463, 165

\bibitem[{{Jenkins}(1952)}]{1952QB813.J45......}
{Jenkins}, L.~F. 1952, {General catalogue of trigonometric stellar
  parallaxes.}, ed. {Jenkins, L.~F.}

\bibitem[{{Judge} {et~al.}(2003){Judge}, {Solomon}, \&
  {Ayres}}]{2003ApJ...593..534J}
{Judge}, P.~G., {Solomon}, S.~C., \& {Ayres}, T.~R. 2003, \apj, 593, 534

\bibitem[{{Linsky} \& {Haisch}(1979)}]{1979ApJ...229L..27L}
{Linsky}, J.~L. \& {Haisch}, B.~M. 1979, \apjl, 229, L27

\bibitem[{Lomb(1976)}]{lomb}
Lomb, N.~R. 1976, Astrophysics and Space Science, 39, 447, 10.1007/BF00648343

\bibitem[{Marquardt(1963)}]{levmar}
Marquardt, D.~W. 1963, 11, 431

\bibitem[{{Mauas} {et~al.}(2012){Mauas}, {Buccino}, {Diaz}, \& {et
  al.}}]{2012arXiv1204.4101M}
{Mauas}, P.~J.~D., {Buccino}, A., {Diaz}, R., \& {et al.} 2012, ArXiv e-prints

\bibitem[{{Micela} \& {Marino}(2003)}]{2003A&A...404..637M}
{Micela}, G. \& {Marino}, A. 2003, \aap, 404, 637

\bibitem[{{Montes} {et~al.}(2001){Montes}, {L{\'o}pez-Santiago}, {G{\'a}lvez},
  \& {et al.}}]{2001MNRAS.328...45M}
{Montes}, D., {L{\'o}pez-Santiago}, J., {G{\'a}lvez}, M.~C., \& {et al.} 2001,
  \mnras, 328, 45

\bibitem[{{Ossendrijver}(2003)}]{2003A&ARv..11..287O}
{Ossendrijver}, M. 2003, \aapr, 11, 287

\bibitem[{{Pallavicini} {et~al.}(1981){Pallavicini}, {Golub}, {Rosner},
  {Vaiana}, {Ayres}, \& {Linsky}}]{1981ApJ...248..279P}
{Pallavicini}, R., {Golub}, L., {Rosner}, R., {Vaiana}, G.~S., {Ayres}, T., \&
  {Linsky}, J.~L. 1981, \apj, 248, 279

\bibitem[{{Pecaut} {et~al.}(2012){Pecaut}, {Mamajek}, \&
  {Bubar}}]{2012ApJ...746..154P}
{Pecaut}, M.~J., {Mamajek}, E.~E., \& {Bubar}, E.~J. 2012, \apj, 746, 154

\bibitem[{{Predehl} \& {Schmitt}(1995)}]{1995A&A...293..889P}
{Predehl}, P. \& {Schmitt}, J.~H.~M.~M. 1995, \aap, 293, 889

\bibitem[{{Robrade} {et~al.}(2005){Robrade}, {Schmitt}, \&
  {Favata}}]{2005A&A...442..315R}
{Robrade}, J., {Schmitt}, J.~H.~M.~M., \& {Favata}, F. 2005, \aap, 442, 315

\bibitem[{{R{\"o}ser} {et~al.}(2008){R{\"o}ser}, {Schilbach}, {Schwan}, , \&
  {et al.}}]{2008A&A...488..401R}
{R{\"o}ser}, S., {Schilbach}, E., {Schwan}, H., , \& {et al.} 2008, \aap, 488,
  401

\bibitem[{{Rosner} {et~al.}(1985){Rosner}, {Golub}, \&
  {Vaiana}}]{1985ARA&A..23..413R}
{Rosner}, R., {Golub}, L., \& {Vaiana}, G.~S. 1985, \araa, 23, 413

\bibitem[{{Scargle}(1982)}]{1982ApJ...263..835S}
{Scargle}, J.~D. 1982, \apj, 263, 835

\bibitem[{{Schlegel} {et~al.}(1998){Schlegel}, {Finkbeiner}, \&
  {Davis}}]{schl98}
{Schlegel}, D.~J., {Finkbeiner}, D.~P., \& {Davis}, M. 1998, \apj, 500, 525

\bibitem[{{Schmitt} \& {Liefke}(2004)}]{2004A&A...417..651S}
{Schmitt}, J.~H.~M.~M. \& {Liefke}, C. 2004, \aap, 417, 651

\bibitem[{{Schwabe}(1843)}]{1843AN.....20..283S}
{Schwabe}, M. 1843, Astronomische Nachrichten, 20, 283

\bibitem[{{Skrutskie} {et~al.}(2006){Skrutskie}, {Cutri}, {Stiening}, \& {et
  al.}}]{2006AJ....131.1163S}
{Skrutskie}, M.~F., {Cutri}, R.~M., {Stiening}, R., \& {et al.} 2006, \aj, 131,
  1163

\bibitem[{{Stauffer} {et~al.}(1994){Stauffer}, {Caillault}, {Gagne}, {Prosser},
  \& {Hartmann}}]{1994ApJS...91..625S}
{Stauffer}, J.~R., {Caillault}, J.-P., {Gagne}, M., {Prosser}, C.~F., \&
  {Hartmann}, L.~W. 1994, \apjs, 91, 625

\bibitem[{{Stern}(1998)}]{1998ASPC..154..223S}
{Stern}, R.~A. 1998, in Astronomical Society of the Pacific Conference Series,
  Vol. 154, Cool Stars, Stellar Systems, and the Sun, ed. R.~A. {Donahue} \&
  J.~A. {Bookbinder}, 223

\bibitem[{Storn \& Price(1997)}]{moncar}
Storn, R. \& Price, K. 1997, Journal of Global Optimization, 11, 341,
  10.1023/A:1008202821328

\bibitem[{{Vaiana} {et~al.}(1981){Vaiana}, {Cassinelli}, {Fabbiano},
  {Giacconi}, {Golub}, {Gorenstein}, {Haisch}, {Harnden}, {Johnson}, {Linsky},
  {Maxson}, {Mewe}, {Rosner}, {Seward}, {Topka}, \&
  {Zwaan}}]{1981ApJ...245..163V}
{Vaiana}, G.~S., {Cassinelli}, J.~P., {Fabbiano}, G., {Giacconi}, R., {Golub},
  L., {Gorenstein}, P., {Haisch}, B.~M., {Harnden}, Jr., F.~R., {Johnson},
  H.~M., {Linsky}, J.~L., {Maxson}, C.~W., {Mewe}, R., {Rosner}, R., {Seward},
  F., {Topka}, K., \& {Zwaan}, C. 1981, \apj, 245, 163

\bibitem[{{Watson} {et~al.}(2009){Watson}, {Schr{\"o}der}, {Fyfe}, \& {et
  al.}}]{2009A&A...493..339W}
{Watson}, M.~G., {Schr{\"o}der}, A.~C., {Fyfe}, D., \& {et al.} 2009, \aap,
  493, 339

\bibitem[{{Wilson}(1968)}]{1968ApJ...153..221W}
{Wilson}, O.~C. 1968, \apj, 153, 221

\bibitem[{{Wilson}(1978)}]{1978ApJ...226..379W}
---. 1978, \apj, 226, 379

\bibitem[{{Wright} {et~al.}(2010){Wright}, {Drake}, \&
  {Civano}}]{2010ApJ...725..480W}
{Wright}, N.~J., {Drake}, J.~J., \& {Civano}, F. 2010, \apj, 725, 480

\bibitem[{Wright {et~al.}(2011)Wright, Drake, Mamajek, \&
  Henry}]{0004-637X-743-1-48}
Wright, N.~J., Drake, J.~J., Mamajek, E.~E., \& Henry, G.~W. 2011, The
  Astrophysical Journal, 743, 48

\bibitem[{{Zacharias} {et~al.}(2004){Zacharias}, {Monet}, {Levine}, \& {et
  al.}}]{2004AAS...205.4815Z}
{Zacharias}, N., {Monet}, D.~G., {Levine}, S.~E., \& {et al.} 2004, in Bulletin
  of the American Astronomical Society, Vol.~36, American Astronomical Society
  Meeting Abstracts, 1418

\end{thebibliography}
\end{document}